\documentstyle[stwol,epsfig]{article}

\def\Journal#1#2#3#4{{#1} {\bf #2}, #3 (#4)}

\def\NPB{{\em Nucl. Phys.} B}
\def\PLB{{\em Phys. Lett.}  B}

\def\ZPC{{\em Z. Phys.} C}

\def\be{\begin{equation}}
\def\ee{\end{equation}}
\def\bea{\begin{eqnarray}}
\def\eea{\end{eqnarray}}
\newcommand{\N}{{\mathcal N}}

\begin{document}

%additional title page for archive

\thispagestyle{empty}
\setcounter{page}{0}

{\large
\twocolumn[
\noindent
%\vspace{-4.0cm}
\rightline{MPI-PhT/96-107}
\rightline{October 1996}
\vspace{2.0cm}
\begin{center}
{\Large \bf QCD Description of Particle Spectra up to LEP-1.5 Energies and
the Running of $\alpha_s$}
\end{center}
\vspace{2.0cm} 
\begin{center}
WOLFGANG OCHS
\end{center}

\medskip 

\begin{center}
 {\it Max-Planck-Institut f\"ur Physik \\
(Werner-Heisenberg-Institut) \\
F\"ohringer Ring 6, D-80805 Munich, Germany} 
\end{center}
\vspace{5.0cm}

\centerline{\bf Abstract}
\bigskip 

\noindent 
Recent results on the energy spectra in QCD jets are reported. Within the
Modified Leading Logarithmic Approximation (MLLA) and the Local Parton Hadron
Duality (LPHD) model one finds a very good description of the $e^+e^-$ data
from the lowest  up to the LEP 1.5 energies.   
A model with fixed $\alpha_s$ can be excluded, already in certain
finite energy intervals. The fits also extrapolate smoothly into the region
of small particle energies, in particular,
the data follow a scaling 
prediction  for the low energy limit derived from the colour
coherence of the soft gluon emission.

\vspace{3.0cm}

\noindent 
to appear in the Proc. of the 28th International Conference on High Energy
Physics,25-31 July 1996, Warsaw, Poland.

\vfill 
]
}
\newpage

\title{QCD DESCRIPTION OF PARTICLE SPECTRA UP TO LEP-1.5 ENERGIES AND THE
RUNNING OF $\alpha_s$}

\author{WOLFGANG OCHS}

\address{Max-Planck-Institute f\"ur Physik (Werner-Heisenberg-Institut)\\
F\"ohringer Ring 6, D-80805 Munich, Germany}


\twocolumn[\maketitle\abstracts
{Recent results on the energy spectra in QCD jets are reported. Within the
Modified Leading Logarithmic Approximation (MLLA) and the Local Parton Hadron 
Duality (LPHD) model one finds a very good description of the $e^+e^-$ data
from the lowest 
up to the LEP 1.5 energies.
A model with fixed $\alpha_s$ can be excluded, already in certain
finite energy intervals. The fits also extrapolate smoothly into the region
of small particle energies, in particular,
the data follow a scaling 
prediction  for the low energy limit derived from the colour
coherence of the soft gluon emission.}]

\section{Introduction}
One of the important predictions of perturbative QCD on the
intrinsic structure of jets concerns the energy spectra of particles.
The particles of low energy $E$ are not multiplied with increasing jet
energy $E_{jet}$ because of colour coherence
in the cascading process and this yields the bell shaped
spectrum in the variable
$\xi = \log \frac{E_{jet}}{E}$, the so-called
 ``hump-backed plateau''.\cite{dfk,bcmm}
%, for reviews, see~\cite{bcm,dkmt1}.

Predictions on the spectrum have been carried out in the 
MLLA which takes into account the leading  double logarithmic (DLA)
results and all next-to-leading corrections of order $\sqrt{\alpha_s}$.
Terms of higher order are included as well, although not completely, which
allow to fulfill the initial condition for the parton cascade at threshold.
These predictions, at the parton level, involve only two parameters: the QCD scale
$\Lambda$ which determines the running of the coupling $\alpha_s$ and the transverse
momentum cut-off $Q_0$ of the gluon emission. Remarkably, the observed
hadron spectra are rather well described by 
the spectrum of partons if
a low value $Q_0 \sim$ 250~MeV of the order of hadronic masses 
is used  and this observation has led to
the LPHD hypothesis \cite{adkt}.
As a justification of this approach is not yet available at a fundamental
level it seems important to determine its range of applicability and its limitations.
Also one would like to know to what extent 
the predictions are sensitive to QCD
as the underlying theory. In this contribution the sensitivity of the energy spectra
to the running $\alpha_s$ and the colour coherence - two properties specific to
QCD as a field theory - are addressed. More details can be found in
refs.~\cite{lo,klo1,klo2}.

\section{Moment Analysis of Particle Spectra}
%\subsection{Analytic results for fixed and running $\alpha_s$}
To analyse the effect of the running $\alpha_s$ it is convenient to work with the
moments of the $\xi$-spectrum of particles which evolve independently of
each other with energy. They are defined by
$<\xi^q(Y, \lambda)> = \int d\xi \xi^q D(\xi,Y,\lambda)/\bar \N$
where we use the logarithmic variables 
$Y=\log(E_{jet}/Q_0)$ and $\lambda=\log(Q_0/\Lambda)$; the normalization is
by the multiplicity $\bar \N$. We consider finally the cumulant moments
$K_q$ which are obtained from the $<\xi^q>$ moments by  
 $K_1 \equiv  <\xi> = \bar \xi$,
 $K_2 \equiv \sigma^2 = <(\xi - \bar \xi)^2>$, $K_3 = <(\xi -
\bar \xi)^3>$, $K_4 = <(\xi - \bar \xi)^4> - 3 \sigma^4$, \dots;
also one introduces the reduced cumulants $\kappa_q \equiv K_q/ \sigma^q$,
in
particular the skewness $s \equiv \kappa_3$ and the kurtosis $k \equiv
\kappa_4$.
The cumulant moments $K_q(Y,\lambda)$ 
behave at high energies like
\bea
K_q(Y,\lambda)& =& K_q(Y_0,\lambda) + \nonumber \\
       & & +\int_{Y_0}^Y dy \left( - \frac{\partial}{\partial \omega} 
\right)^q \gamma_{\omega}[\alpha_s(y,\lambda)] \biggl|_{\omega=0}
\label{mom2}                                      
\eea
where $Y_0$ is the initial energy. Here $\gamma_{\omega}[\alpha_s(y,\lambda)]$
denotes the anomalous
dimension of the Laplace transform of the $\xi$-spectrum 
$dn/d\xi\equiv D(\xi,Y,\lambda)$.
Eq.(\ref{mom2}) 
shows directly the sensitivity of the cumulant moments to the
running of $\alpha_s$. In particular, for fixed coupling $K_q \sim Y$
at high energies.
In our applications, we take into account the initial condition at
threshold for the jet to consist of only one parton 
which implies $K_q=0$ at $Y_0=0$.

The MLLA prediction for the $\xi^q$ moments 
can be written for arbitrary $Q_0$ and $\Lambda$ as~\cite{dktint}: 
\be
< \xi^q > = \frac{1}{\bar \N} \sum_{k=0}^q {q \choose k} ( N_1 L_k^{(q)}
+ N_2 R_k^{(q)} )
\label{moments:full}                              
\ee                                               
where $N_1$, $N_2$, $L_k^{(q)}$ and $R_k^{(q)}$ are known functions of
a = 11$N_c$/3+2$n_f$/3$N_c^2$, b $\equiv$ (11$N_c$-2$n_f$)/3, and
$\lambda$ where $n_f,N_C$ denote the numbers of flavours and colours. 
Comparing these formulae to the experimental data on moments,
\cite{lo} best agreement is found for the ``limiting spectrum'', where
the two parameters  coincide, i.e., $Q_0$ = $\Lambda$, or $\lambda$ = 0.
In this case the formulae simplify and
the moments can be expressed\cite{dktint} in terms of
the parameter $B \equiv a/b$                          
and the variable $z \equiv \sqrt{ 16 N_c Y / b }$ as:
\begin{eqnarray}                                               
\frac{<\xi^q>}{Y^q}& =& P_0^{(q)}(B+1,B+2,z) + \nonumber \\
    & & +\frac{2}{z}
\frac{I_{B+2}(z)}{I_{B+1}(z)} P_1^{(q)}(B+1,B+2,z)\nonumber\\
\label{moments:LS}                                
\end{eqnarray}                                               
where $P_0^{(q)}$ and $P_1^{(q)}$ are polynomials 
of order \\
$2(q-1)$ in $z$.

These expressions extrapolate smoothly to threshold, where they are determined by
the initial condition for a single parton. Similarly one can derive the
complete MLLA results for fixed $\alpha_s$
\cite{lo}, which simplify for high energies to
\be 
\bar \xi_{fix}  =   \left[ 1 +  
\frac{\eta }{\bar \gamma_0 }  \right] \frac{Y}{2}
\quad , \quad 
\sigma^2_{fix} = \frac{\gamma_0^2}{4 \bar \gamma_0^3} Y
\label{momfix}
\ee
\be
s_{fix} = - \frac{3 \eta}{\gamma_0} \frac{1}{\sqrt{\bar \gamma_0 Y}}
\quad , \quad 
k_{fix} = \frac{3 ( 4 \eta^2 - \gamma_0^2)}{\gamma_0^2 \bar \gamma_0}
\frac{1}{Y}
\nonumber
\ee
where $\eta = a \gamma_0^2/8N_C$, $\bar \gamma_0 \equiv \sqrt{\gamma_0^2 +
\eta^2}$ and $\gamma_0 = \sqrt{2 N_C \alpha_s/\pi}$.

\section{Relating parton and hadron spectra}

Strictly speaking the QCD results are only reliable for $E \gg Q_0$.
In order to determine the moments one has to integrate the spectra over the
full range and to extrapolate to the soft limit.  
The experimental particle densities are usually
given as function of momenta.
The same kinematic limits of parton and hadron spectra is obtained if 
the same effective mass $Q_0$ is assigned to the
charged particles which also limits the
energy of the partons ($E_h = E_p \ge Q_0$).

Furthermore, a common behaviour of the parton and hadron spectra
near the boundary  
can be obtained if the spectra are related by \cite{dfk,lo}
\be
E_h \frac{dn(\xi_E)}{dp_h}=K_h E_p\frac{dn(\xi_E)}{dp_p} 
\equiv K_h D(\xi_E,Y,\lambda) 
\label{dual}
\ee
at the same energy $E$ or $\xi_E \equiv \xi = \log (E_{jet}/E)$, where
$E_h = \sqrt{p_h^2+Q^2_0}$ and $E_p = p_p$; $K_h$ denotes a normalization
parameter.
With this choice both spectra
vanish linearly in $\xi$; this corresponds to  the invariant
distribution of hadrons 
$E \frac{dn}{d^3p}$ approaching a constant value in the soft limit
$p \to 0$, as suggested  by the experimental data.
The prescription~(\ref{dual}) 
represents a minimal extension of the QCD prediction for $E \gg Q_0$ 
towards the boundary at $E = Q_0$ but is not unique;
other prescriptions using a smaller value of $Q_0\sim m_\pi$ have also been
applied  (see discussion in \cite{klo2}). 
  
  %%%%%
\begin{figure}
          \begin{center}
\mbox{\epsfig{file=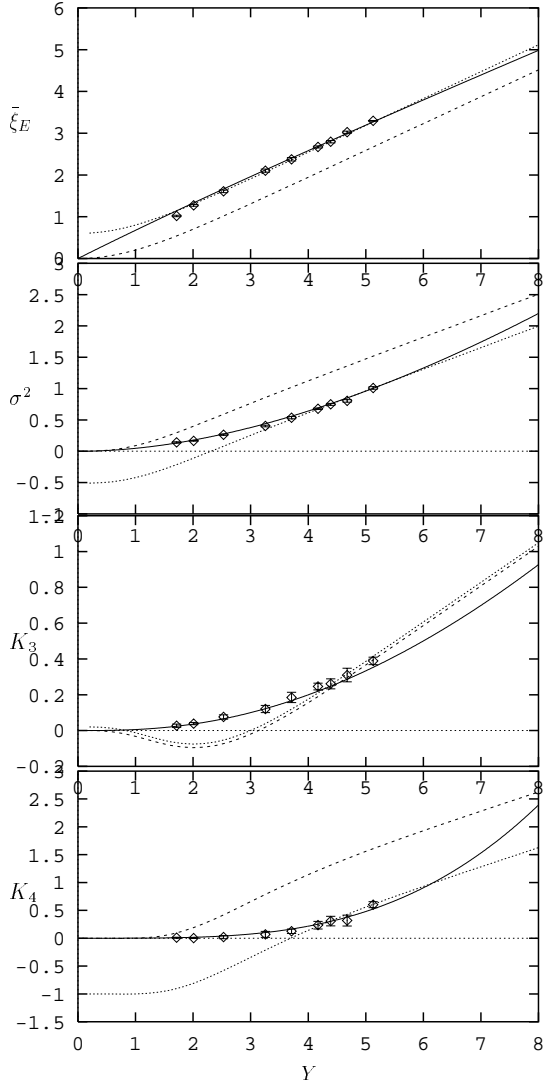,bbllx=4.5cm,bblly=4.cm,bburx=20.cm,
bbury=26.cm,height=14.cm}}
       \end{center}
\caption{The first four cumulant moments $K^q$ of the  
charged particles' energy spectra $E dn/dp$ vs. $\xi_E$, 
shown as a function of $Y$ 
for $Q_0$ = 270 MeV.   
Predictions of the Limiting Spectrum of MLLA with running $\alpha_s$ (solid
lines), of MLLA with fixed $\alpha_s$ (= 0.214) (dashed lines) and of MLLA 
with fixed $\alpha_s$ normalized by hand to the experimental point 
at $\protect\sqrt{s}$ = 44 GeV (dotted lines) are also shown; in all cases $n_f$
 = 3 (from Ref. 4).} 
\label{moments}
\end{figure}

\section{Comparison with experimental data}
The moments $<\xi^q>$ are determined
\cite{lo} from the spectra $E dn/dp$ vs. $\xi_E$ after appropriate transformation
of the measured $x_p = 2p/\sqrt{s}$ spectra of charged particles using
$E = \sqrt{p^2+Q_0^2}$ and therefore depend on $Q_0$. First a fit of the two
parameters $Q_0$ and $\Lambda $ is obtained
by comparing the moments for a selected $Q_0$ with
the theoretical predictions from Eq.(\ref{moments:full}) 
for different $\Lambda $. The best agreement
with the data was obtained for $Q_0 \approx \Lambda \approx  (270 \pm 20)$ MeV.
It was not possible to obtain a satisfying description for the fixed $\alpha_s $ case.
Choosing $\gamma_0$ = 0.64 and the same $Q_0$ = 270~MeV a good description of the
multiplicity $\bar \N$ 
and the slope of $\bar \xi$ vs. $Y$ could be obtained but not for the other
quantities. In Fig. 1 we show the evolution of the first cumulant moments 
$K_q$ with
enery for running $\alpha_s $ and fixed $\alpha_s $. Note that these predictions
depend only on the two parameters $Q_0 $ and $\Lambda $ which actually coincide
in the fit. The absolute normalization of the moments
is given at threshold $Y_0$=0 by $K_q$=0.

As this is an application of perturbative QCD to an extreme limit 
one could make the
weaker (more conventional) assumption and choose a higher starting energy, say
    $Y_0 = 2$ $(\sqrt{s} \approx 4 $ GeV).
In this case for each moment an extra free parameter $K_q(Y_0)$ had to be introduced
which, however, 
 would not improve the fit for running $\alpha_s $ essentially; 
a backward evolution from $Y_0$ would again reproduce 
approximately the initial
condition at threshold. Therefore, the initial condition yields the highly
constrained fit with only two parameters (the normalization parameter
$K_h$ only enters the multiplicity $\bar \N$, not shown here).

In case of fixed $\alpha_s $ one may ask the same question, whether the fit could be
improved by choosing a higher starting value $Y_0$, i.e. to assume a 
transition from
a running coupling to a fixed coupling regime at $Y_0$
(This would correspond to a parallel shift of the dashed curve to the data
at $Y_0$).
As can be seen in Fig. 1 the energy dependence of the fixed and running $\alpha_s $
curves is quite different so that an improvement cannot be obtained in a finite
energy interval. Rather, such a procedure can be excluded for a $Y$ interval
of about 1-2 units.
An example is shown by the dotted curve which corresponds to
$Y_0 = 4.3$ $ (\sqrt{s}$ = 44~GeV).
The trend at yet larger energies is again different, in particular for the
higher moments.

The dependence on the number of flavours $n_f$ has been studied as well
\cite{lo,klo1}. As the jet evolution is dominated at all energies by the
gluon emission at the lower transverse momenta, the choice $n_f$ = 3 is a good
approximation up to LEP energies.

\begin{figure}
          \begin{center}
\mbox{\epsfig{file=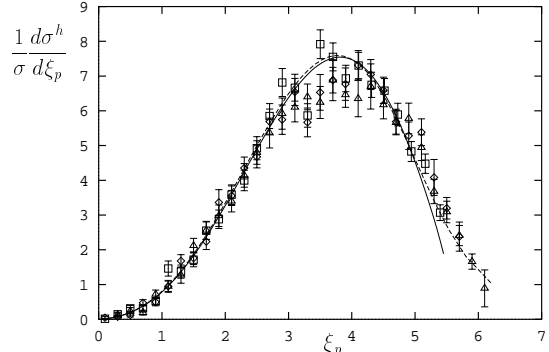,bbllx=3.cm,bblly=18.5cm,bburx=22.5cm,bbury=27.cm,width=10cm}}
       \end{center}
\caption{Charged particle inclusive momentum distribution at 
LEP-1.5 from ALEPH (diamonds), 
DELPHI (squares) and OPAL Collaborations
(triangles) in comparison with  theoretical predictions of the 
Limiting Spectrum with $Q_0$ = 270 MeV (solid line). Dashed 
lines show the predictions of the Limiting Spectrum after  correction for 
kinematical effects (from Ref. 5).}
\label{shape}
\end{figure}

The recent LEP-1.5 data are also well accounted for
\cite{klo1} by the same theoretical scheme as is shown in Fig. 2. Here the
experimental $\xi$-spectrum is shown in comparison with the limiting spectrum
(setting $\xi_p = \xi_E$) which terminates at $Y \sim$ 5.3 and
after rescaling $\xi_E \to \xi_p$ with relation (\ref{dual}) which takes into account
the boundary effects. Also the moments of the distribution and the position
of the maximum are well reproduced by the theoretical predictions 
at this energy.

\begin{figure}
          \begin{center}
\mbox{\epsfig{file=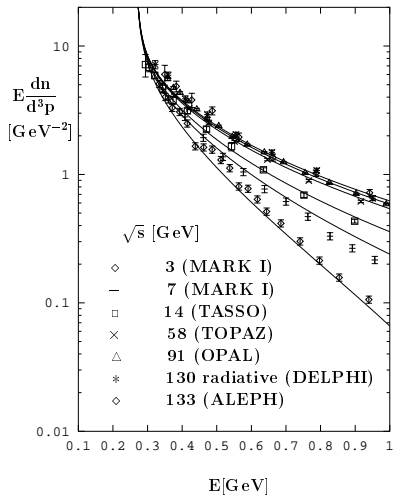,width=.4\linewidth,bbllx=5.5cm,bblly=10.5cm,bburx=13.5cm,bbury=26.5cm}}
          \end{center}
\caption{Invariant density $E dn/d^3p$ of charged particles 
in $e^+e^-$ annihilation 
as a function of the particle energy $E = \protect\sqrt{p^2 + Q_0^2}$ 
at $Q_0$ = 270 MeV at various $cms$ energies 
in comparison with the MLLA predictions % (normalization $K_h$ = 0.45), 
(from Ref. 6).}
\end{figure}

\section{Soft limit of energy spectrum}
Having observed the good fit of the QCD-LPHD prediction 
not only towards high but also towards low $cms$
energies $\sqrt{s}$ we now turn to the behaviour of the fits at low particle
energy $E$. As the emission of very soft gluons from all other partons in the
jet is coherent the production rate of such particles is nearly energy independent
\cite{adkt}. Indeed, the analytic results in the DLA and MLLA show the scaling
behaviour of the spectrum in the soft limit and the production rate is essentially
determined by the colour charge of the initial partons
\cite{klo2}. Remarkably, the observed hadrons follow the trend of this prediction
as can be seen in Fig. 3 which shows the invariant particle 
density as a function
of particle energy $E$ (assuming the effective 
charged particle mass $Q_0$ = 270~MeV).
The data from all energies in $e^+e^-$ annihilation approach a common limit 
for $E \to Q_0$ $(p \to 0)$. The curves which represent the MLLA
calculation show again the scaling limit and reproduce the
dependence on energy $E$ and $cms$-energy $\sqrt{s}$ rather well.
This suggests that LPHD persists towards very low particle energies.
Various further tests of this picture have been proposed
\cite{klo2} by studying the dependence of the soft limit $I_0$ on
the colour charge of the primary partons in $e^+e^- \to$ 3 jets,
deep inelastic scattering processes and hadronic collisions.

\section{Conclusions}
The particle energy spectra are well described by the QCD parton shower
in the MLLA assuming a close similarity between partons and hadrons
according to the LPHD approach. This similarity appears to work 
not only for the highest but also down to
rather low $cms$ energies as well as to low particle energies.
This agreement is by no means trivial. For example, replacing the running
$\alpha_s$ by the fixed $\alpha_s$ yields predictions quite incompatible
 with experiment, also if the threshold for the onset of a fixed $\alpha_s$
regime is increased. The scaling prediction derived from colour coherence for
the soft particles works well for hadrons.

This may suggest \cite{lo} to view the final stage of hadron production
through resonances, as explicitly incorporated into Monte Carlo hadronization
models, as being dual to the partonic evolution with running coupling
down to the low scale of a few hundred MeV.

\section*{Acknowledgements}
I would like to thank Valery Khoze and Sergio Lupia for many discussions
and the collaboration on the subjects of this talk.

\end{document}